\documentclass{iaus}
\usepackage{graphicx}
  \checkfont{eurm10}
  \iffontfound
    \IfFileExists{upmath.sty}
      {\typeout{^^JFound AMS Euler Roman fonts on the system,
                   using the 'upmath' package.^^J}%
       \usepackage{upmath}}
      {\typeout{^^JFound AMS Euler Roman fonts on the system, but you
                   dont seem to have the}%
       \typeout{'upmath' package installed. iaus.cls can take advantage
                 of these fonts,^^Jif you use 'upmath' package.^^J}%
      }
  \else
  \fi

  \checkfont{msam10}
  \iffontfound
    \IfFileExists{amssymb.sty}
      {\typeout{^^JFound AMS Symbol fonts on the system, using the
                'amssymb' package.^^J}%
       \usepackage{amssymb}%

      }{}
  \fi

  \IfFileExists{amsbsy.sty}
    {\typeout{^^JFound the 'amsbsy' package on the system, using it.^^J}%
     \usepackage{amsbsy}}
    {}

\newsavebox{\astrutbox}
\sbox{\astrutbox}{\rule[-5pt]{0pt}{20pt}}

\newcommand\etal{\mbox{\textit{et al.}}}

\newcommand{\msun}{$M_{\odot}$}

\newcommand{\sig}{$\sigma$}
\newcommand{\mbh}{$M_{\rm BH}$}
\newcommand{\chandra}{{\it Chandra}}

\title[The Interplay among Black Holes, Stars and ISM in Galactic 
       Nuclei]{Black Hole Growth \& the \mbh--Bulge Relations}

\author[S. Mathur]%
{Smita Mathur \& Dirk Grupe }

\affiliation{Astronomy Department, The Ohio State University, 140 West 18th Ave., Columbus, OH 43210, USA. email: smita@astronomy.ohio-state.edu}

\pubyear{2004}
\volume{222}
\pagerange{1--8}
\date{?? and in revised form ??}
\jname{The Interplay among Black Holes, Stars and ISM \\in Galactic Nuclei}
\editors{Th. Storchi Bergmann, L.C. Ho \& H.R. Schmitt, eds.}
\begin{document}

\maketitle

\begin{abstract}
We present the black hole mass--bulge velocity dispersion relation for a
complete sample of 75 soft X-ray selected AGNs. We find that the AGNs with
highest accretion rates relative to Eddington lie below the \mbh--\sig\
relation of broad line Seyfert 1s, confirming the Mathur \etal\ (2001)
result. The statistical result is robust and not due to any systematic
measurement error. This has important consequences towards our
understanding of black hole formation and growth: black holes grow by
accretion in well formed bulges. As they
grow, they get closer to the \mbh--\sig\ relation for normal
galaxies. The accretion is highest in the beginning and dwindles as 
time goes by. Our result does not support theories of the \mbh--\sig\
relation in which the black hole mass is a constant fraction of the
bulge mass/ velocity dispersion {\it at all times} or those in which
bulge growth is controlled by AGN feedback.
\end{abstract}

\firstsection % if your document starts with a section,
              % remove some space above using this command.
\section{Introduction}

The observation of a tight correlation between the velocity dispersion
\sig\ of the  bulge in a galaxy and the mass of its nuclear
black-hole \mbh\ was a surprising discovery over the last few years
(\cite{geb00a, ferr00, merr01})). Even more surprisingly, the above
relation for normal galaxies was also found to extend to active galaxies
(\cite{geb00b, ferr01}). A lot of theoretical models provide explanations
for the \mbh--\sig\ relation in the framework of models of galaxy
formation, black hole growth and the accretion history of active
galactic nuclei (\cite{haeh03, haeh98}, \cite{king03}). To
understand the origin of the \mbh--\sig\ relation, and to discriminate
among the models, it is of interest to follow the tracks of AGNs on the
\mbh -- $\sigma$ plane.

Mathur \etal\ (2001) suggested that the narrow line Seyfert 1 galaxies
(NLS1s), a subclass of Seyfert galaxies believed to be accreting at a
high Eddington rate, do not follow the \mbh--\sig\ relation. Here we
present our results based on a complete sample of 75 soft X-ray
selected AGN.

Note also that NLS1s are interesting objects as they occupy one extreme
end of the ``eigenvector 1'' relation of AGNs (\cite{bor03}). The most
widely accepted paradigm for NLS1s is that they accrete at close to the
Eddington rate and have smaller black hole masses for a given luminosity
compared to BLS1s. Finding their locus on the \mbh--\sig\ plane is
therefore a worthwhile experiment anyway as we will either find that
they occupy a distinct region compared to BLS1s or that they don't. The
first option is interesting for our understanding of black hole
growth. On the other hand if we find that NLS1s follow the \mbh--\sig\
relation like the BLS1s, it has important implications towards our
understanding of the AGN phenomenon. We already have good evidence for
smaller BH masses of NLS1s, at a fixed luminosity. If they follow the
\mbh--\sig\ relation, it would imply that NLS1s preferentially reside in
galaxies with bulges of smaller velocity dispersion. This would be
direct evidence for the dependence of AGN properties on their large
scale galactic environment.

\vspace*{-0.5cm}
\section{The \mbh--\sig\ relation}

We use luminosity and FWHM(H$\beta$) as surrogates for black hole mass
and FWHM([OIII]) as a surrogate for the bulge velocity dispersion. See
Grupe \& Mathur (2004; Paper I here on-wards) for the details of sample
selection and for the validity of our method to estimate \mbh\ and \sig. 

Figure 1 shows that BLS1s and NLS1s occupy two distinct regions in the
\mbh -- \sig~ plane.  For a given velocity dispersion NLS1s tend to show
smaller smaller black hole masses than BLS1s. If true, this is an
important result. We emphasize that this is a statistical result; errors
on both \mbh\ and \sig\ are large (figure 1).  As discussed in Paper I,
this is not due to systematically underestimating BH masses of
NLS1s. Moreover, BH mass estimates using two completely different
methods give the same result: in Mathur \etal\ 2001, \mbh\ was
determined by fitting accretion disk models to SEDs and in Czerny \etal\
2001, power-spectrum analysis was used.

\begin{figure}
 \vspace*{-2.0cm} 
 \includegraphics[width=5cm]{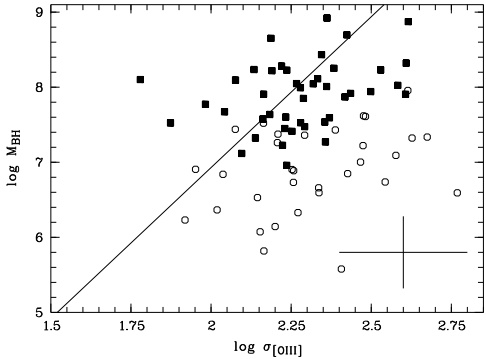} 
 \caption{Velocity dispersion $\sigma_{\rm [OIII]}$ vs. log $M_{\rm
 BH}$(H$\beta$). NLS1s are marked as open circles and BLS1s as filled
 squares.  Black hole masses are given in units of \msun.  The solid
 line marks the relation of \cite{tre02}. The cross at the bottom right
 hand corner represents a typical error bar.}
\end{figure}

In Paper I we also scrutinize the use of FWHM([OIII]) as a surrogate for
the bulge velocity dispersion. Clearly, there is a large scatter in the
$\sigma_{[OIII]}-\sigma_*$ relation. The important thing to note,
however, is that there is no systematic {\it difference} for the two
classes BLS1s and NLS1s. We also explore possible problems specific to
NLS1s, like strong FeII emission and [OIII] asymmetry, and find these do
not affect the result either. We thus conclude that BLS1s and NLS1s
occupy distinct regions in the \mbh -- \sig~ plane. This is clearly seen
from figure 2 which plots the cumulative distribution of the ratio
log(\mbh/\sig) for BLS1s and NLS1s. The two classes are significantly
different, with formal Kolmogorov-Smirnov (K-S) test probability of
being drawn from the same population $< 0.001$.

\begin{figure}
 \vspace*{-2.0cm}
 \hspace*{-1.0cm}
 \includegraphics[width=8cm]{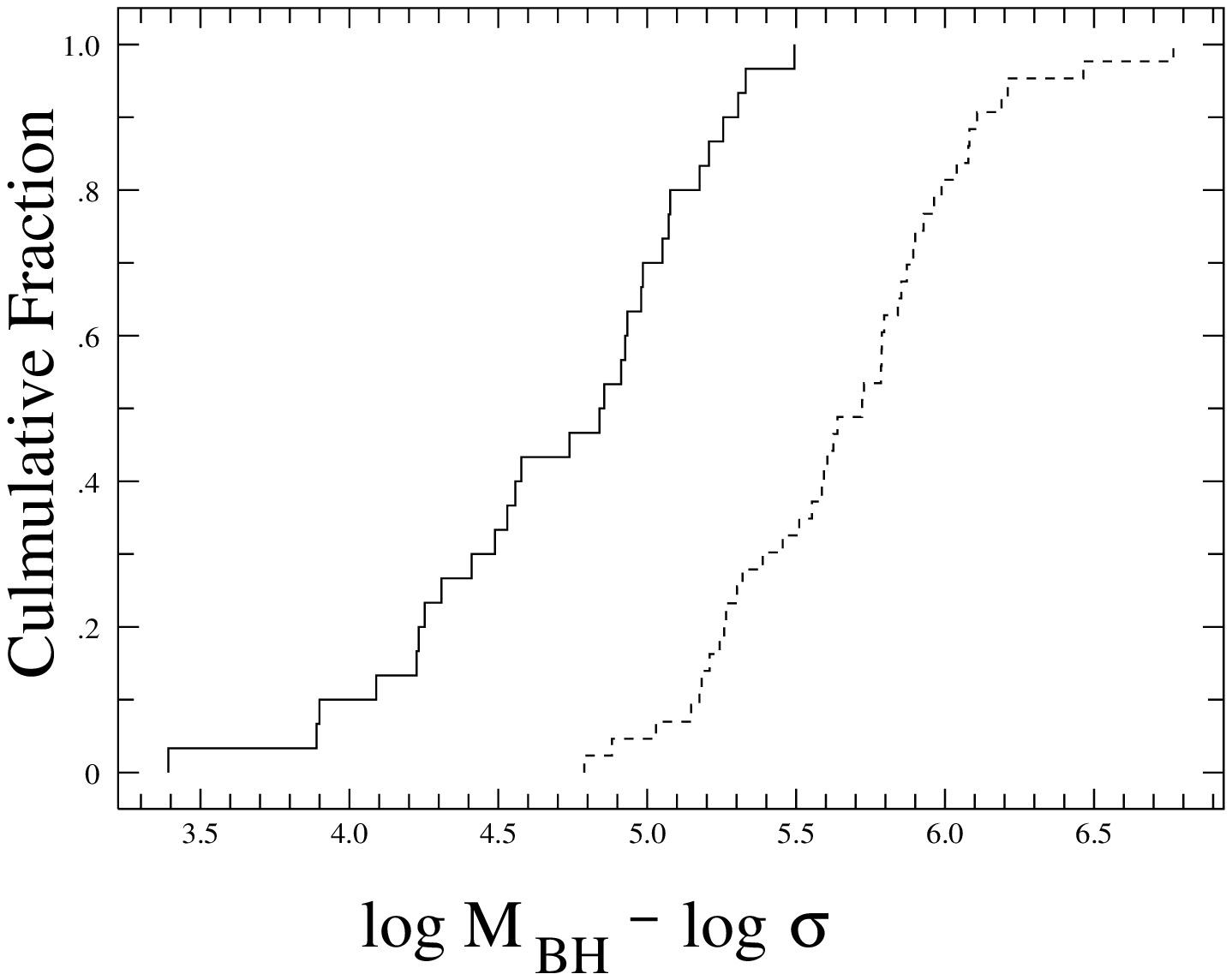} 
 \hspace*{-1.0cm}
 \includegraphics[width=8cm]{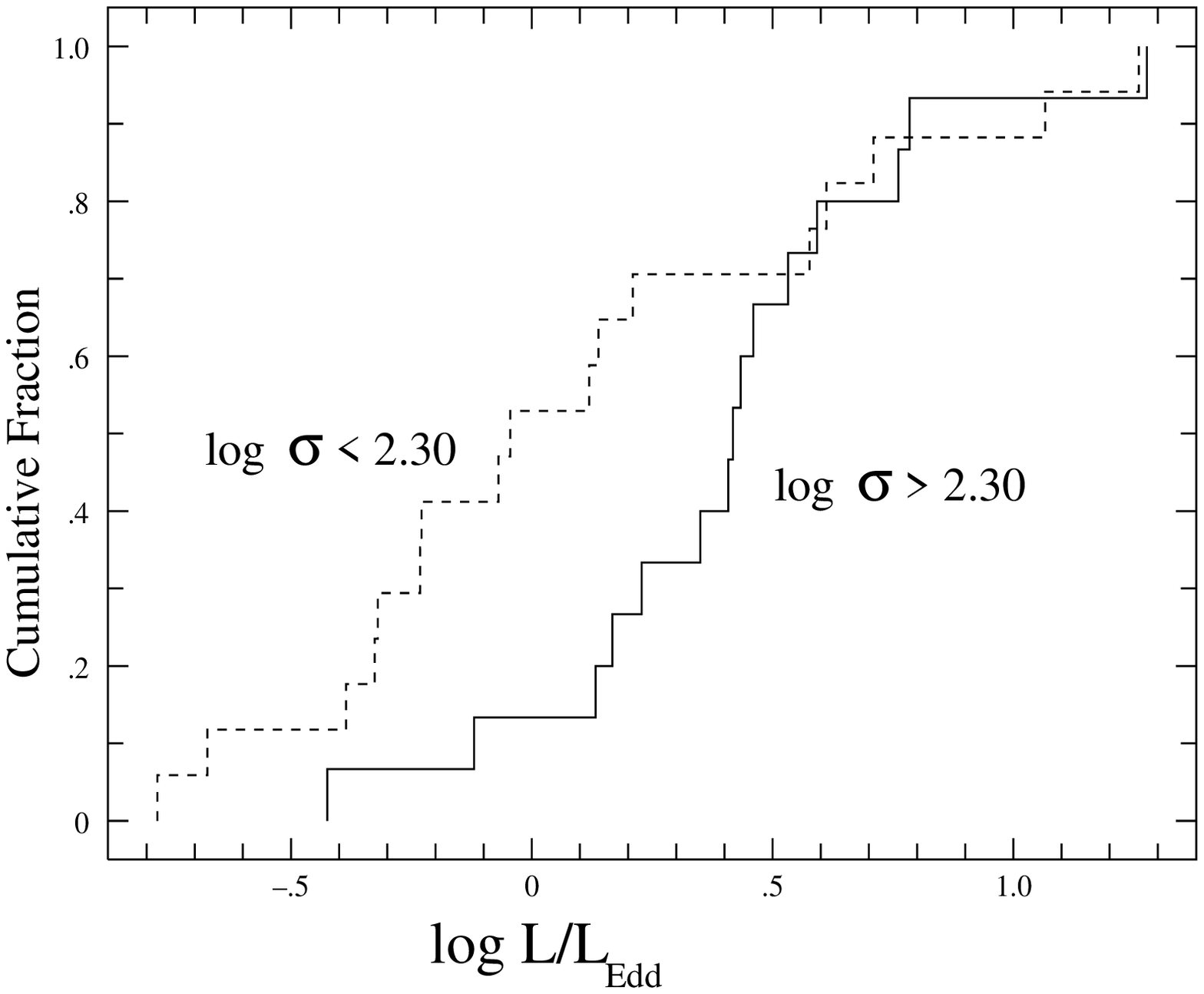}
 \vspace*{-2.5cm}
 \caption{{\bf Left} Cumulative fraction for a KS test of the
 distributions of the black hole mass $M_{\rm BH}$ divided by the
 stellar velocity dispersion $\sigma$.  The distribution of NLS1s is
 shown as a solid line and BLS1s are shown as a dashed line. The two
 populations are clearly different. }
 \caption{{\bf Right} Cumulative fraction for a KS test of the
 distributions of ${\rm L/L_{Eddington}}$ for the two subsamples of
 NLS1s. The large \sig\ sources also appear to be with large ${\rm
 L/L_{Eddington}}$ }
\end{figure}

In Paper I, we interpret this result in terms of black hole growth:
black holes grow significantly by accretion in well formed bulges and
they reach the \mbh--\sig\ relation eventually as the growth is
complete. This scenario is consistent with the models of
Miralda-Escud\'e \& Kollmeier (2004) and also with the suggestion of
NLS1s being young AGNs (Mathur 2000). While our statistical result is
robust, the same is not obvious about its interpretation. This is
because some NLS1s lie on/ very close to the \mbh--\sig\ relation
(figure 1, Mathur \etal\ 2001, \cite{bian03}). Two NLS1s in Ferrarese
\etal\ (2001) also lie on \mbh--\sig\ relation. If we are to interpret
the observations in terms of black hole growth by the highly accreting
NLS1s, why have some NLS1s already reached their ``final'' mass?

The first hint towards the resolution of the above conflict came from
the observations of Williams, Mathur \& Pogge (2004). In \chandra\
observations of NLS1s, they found a significant fraction with flat X-ray
spectra, and with low accretion rate relative to Eddington (\.{m}). In
the framework of the black hole growth scenario, such objects may then
be the ones close to the \mbh--\sig\ relation, as they would have
already gone through their high \.{m} state and their black holes
accumulated most their mass.

To test this hypothesis, we divided our NLS1 sample in two parts, with
low and high values of \sig\, with a boundary at log$\sigma_{\rm
[OIII]}$=2.3. The choice of the boundary came from visual inspection
of figure 1, where it appears that the NLS1s with log$\sigma_{\rm
[OIII]}$ below this value tended to be much closer to the \mbh--\sig\
relation. Figure 3 compares the distribution of ${\rm L/L_{Eddington}}$
for the two samples. The K-S cumulative distribution for the two samples is
significantly different with the formal K-S test probability of being
drawn from the same population P=0.05. Consistently, we also find that
the low \sig\ NLS1s have flatter X-ray spectra.

%\begin{figure}
% \vspace*{-2.0cm}
% \includegraphics[width=8cm]{fig4.ps} 
% \vspace*{-2.0cm}
% \caption{Cumulative fraction for a KS test of the distributions of
% X-ray power-law slope $\alpha_X$ for the two subsamples of NLS1s. The
% large \sig\ sources also appear to be with large $\alpha_X$}
%\end{figure}

The above results show that NLS1s are a mixed bag, some with steep
$\alpha_X$ but some with flat and some with large \.{m} and some with
small. The objects with large \.{m} are the ones which lie below the
\mbh--\sig\ relation of dead black holes and black holes with low \.{m}. 
Thus the interpretation presented in Paper I appears to be secure: black
holes grow in mass substantially in their highly accreting phase. As
they grow, they approach the \mbh--\sig\ relation for normal
galaxies. The mass growth in a low accretion phase, as in BLS1s and also
in some NLS1s, appears to be insignificant. Any theoretical model
attempting to explain the \mbh--\sig\ relation will have to explain the
above observations.

\vspace*{-0.5cm}
\section{Further Tests}

Needless to say, it is vital to measure \mbh\ and \sig\ accurately to
confirm the above result. Black hole mass estimates based on H$\beta$
widths are quite secure, but the same cannot be said about estimates of
\sig\ based on [OIII] widths. {\bf Even if FWHM([OIII]) is not a good
surrogate for \sig\, the nature of our result is such that
$\sigma_{[OIII]}-\sigma$ would then have to be {\it different} for BLS1s
and NLS1s}, and this is most likely not the case as discussed in Paper
I. Moreover, there is no observational result to support such a
difference. If NLS1s had larger outflows, then they could have disturbed
their narrow-line regions more compared to BLS1s. Again, there are no
observations supporting such a case; on the contrary, absorbing outflows
are seen less often in NLS1s (Leighly 1999). Larger ${\rm
L/L_{Eddington}}$ in NLS1s does not necessarily imply larger effective
radiation pressure. On the contrary, in objects with large soft X-ray
excesses, like NLS1s, the {\it absorbed} radiation is actually much
smaller (Morales \& Fabian 2002). There is also general lore that highly
accreting sources with large \.{m} should have large outflows. While low
efficiency accretion must lead to outflows (as in ADIOS, Blandford \&
Begelman 1999), the same is not true for efficient accretion as in
bright Seyferts and quasars. Large outflows are observed in highly
accreting sources like broad absorption line quasars, but that depends
upon the ratio of gas supply to Eddington accretion rate, and is not
inherent to the accretion process itself (R. Blandford, private
communication).

Bulge velocity dispersion is usually measured with CaII triplet line and
this technique has been used to measure \sig\ in two NLS1s (Ferrarese \etal\
2001). However, for many of the NLS1s in our sample, the CaII lines fall
in the water vapor band in the Earth's atmosphere. In many NLS1s for
which CaII line is accessible from ground, CaII is observed in emission
rather than in absorption (Rodriguez-Ardila \etal 2002). This makes 
use of CaII absorption features to determine \sig\ difficult for the
targets of interest. We plan to use two different methods for
alternative estimates of \sig: (1) use of the CO absorption band-head at
2.29 microns to measure \sig\ directly; and (2) high resolution imaging
of the NLS1 host galaxies to measure surface brightness distribution of
bulges. One can then use the fundamental plane relation to determine
\sig. Alternatively, we will determine the bulge luminosities and find
the locus of NLS1s on the \mbh-L$_{Bulge}$ relation.  Once again, the
objective is to find out if there exists a statistical difference in the
ratio of black hole mass to bulge luminosity for the two populations of
BLS1s and NLS1s. We plan to use all these methods to fully understand
the role of accretion in black hole growth, and to determine the locus
of highly accreting AGNs on the \mbh--bulge relations.

\end{document}